\documentclass[aps,prb,preprint,onecolumn,superscriptaddress]{revtex4-2}
\usepackage{amssymb,amsmath,amsfonts}
\usepackage{booktabs}
\usepackage{graphicx} 
\usepackage{float} 
\usepackage{natbib} 
\usepackage{color,soul}
\usepackage{xcolor}
\usepackage[colorlinks=true,bookmarks=false,citecolor=blue,urlcolor=blue]{hyperref}
\usepackage{array}
\usepackage{lmodern} 
\usepackage{soul}

\graphicspath{ {./Figures/} }

\newcommand{\w}{\omega }

\renewcommand{\Re}{\frak{R}\mathrm{e} }
\renewcommand{\Im}{\frak{I}\mathrm{m} }

\newcommand{\eps}{\varepsilon}

\newcommand{\bg}{ \mathrm{bg} }

\newcommand{\qingdao}{
Qingdao Innovation and Development Center of Harbin Engineering University, Qingdao 266000, Shandong, China
}

\newcommand{\itmo}{ITMO University, St. Petersburg 197101, Russia}

\begin{document}

\title{Polaritonic spectra of optical Mie voids}

\author{Evgeny Ryabkov}
\affiliation{\qingdao}
\affiliation{Center for Photonics and 2D Materials, Moscow Institute of Physics and Technology, Dolgoprudny 141700, Russia}

\author{Mingzhao Song}
\email[]{kevinsmz@foxmail.com}
\affiliation{\qingdao}

\author{Andrey A. Bogdanov}
\email[]{a.bogdanov@hrbeu.edu.cn}
\affiliation{\qingdao}
\affiliation{\itmo}

\author{Denis G. Baranov}
\email{denis.baranov@phystech.edu}
\affiliation{Center for Photonics and 2D Materials, Moscow Institute of Physics and Technology, Dolgoprudny 141700, Russia}

\begin{abstract}
The progress in understanding the optical and microscopic properties of polaritons relies on various optical cavities to confine electromagnetic radiation, which causes a demand for new platforms with higher $Q$-factors and better fabrication robustness.
In this context, so called Mie voids -- spherical cavities inside a dielectric medium, where the light confinement occurs due to refractive index contrast at the air-dielectric interface -- present a substantial interest.
Here, we theoretically study the resonant characteristics and polaritonic spectra of spherical Mie cavities loaded with resonant media, as well as address the inverted problem, where a Mie void is formed inside a resonant dispersive medium.
We establish approximate analytical expressions for the $Q$-factors of Mie void cavities,
find the parameter ranges of spherical voids leading to the regimes of weak and strong light-matter coupling and analyze the concomitant effects,
such as $Q$-factor enhancement and spatial field localization, from the polaritonic perspective. 
Our result could be valuable for the design of new polaritonic systems.
\end{abstract}


\maketitle
\newpage

\section{Introduction}

Polaritons are hybrid eigenstates of an optical cavity interacting with material excitations -- such as an electronic or a vibrational transitions -- in the non-perturbative regime \cite{Khitrova2006, Torma2015}.
Thanks to their hybrid composition, polaritons offer opportunities to modify various microscopic properties of matter \cite{Ebbesen2016, jarc2023cavity}, and even to control the rates of certain chemical reactions by altering molecular energy levels
\cite{Hutchison2012, Galego2016, Herrera2016, Thomas2019, schafer2019modification, YuenZhou2022}.
The study of polaritons in the optical and infrared domains relies on various optical cavities to confine electromagnetic radiation. 
The progress in this field causes a steady demand for new platforms with higher $Q$-factors, easier fabrication requirements, and better fabrication robustness \cite{Baranov2018}.

Frequently used types of optical cavities incorporated in polaritonic systems include traditional Fabry\textendash P\'erot (FP) cavities \cite{thompson1992observation, Lidzey1998, Reithmaier2004}, photonic crystal cavities \cite{Yoshle2004}, propagating surface modes \cite{Bellessa2004, Hakala2009}, individual plasmonic nanocavities \cite{Savasta2010,Zengin2015, Chikkaraddy2016, Santhosh2016}, lattice resonances \cite{Vakevainen2014}. 
More recent research efforts resort to bulk \cite{Menghrajani2020} or so-called cavity-free polaritons \cite{Canales2021,thomas2021cavity, tserkezis2024self, canales2024self}, where the confinement of electromagnetic radiation is facilitated by the same material, which hosts the material excitation itself \cite{Platts2009}.

In this context, voids inside a dielectric medium, where light confinement occurs due to the refractive index contrast at the air-dielectric interface, present a substantial interest. Depending on their geometry and dielectric properties of the host materials, such cavities were theoretically and experimentally proven to be a versatile platform for scattering control \cite{Mundy1974, Vanecek1991, Shalin2010, Panda2020} and absorption enhancement \cite{Mann2011}, as well as general resonance tailoring \cite{Chen1998, Sarbajna2024}. Recently, spherical void resonators embedded in a dielectric medium have been reintroduced by Hentschel et al. \cite{hentschel2023dielectric} as dielectric Mie voids, since their modes can be described by Mie theory \cite{Mie1908}. 
In a way, one can think of these optical modes as Mie resonances "inside-out" \cite{Kuznetsov2012, Evlyukhin2012, Koshelev2021}.

Despite the extensive examination and utilization of Mie voids, they have not yet been considered as platforms for realizing strong light-matter coupling. Polaritonic solutions in the inverted cases --  dielectric spheres incorporating resonant media -- have been extensively studied in recent works as self-hybridized polaritons \cite{tserkezis2018mie, todisco2020magnetic, tserkezis2024self}. In this regard, the question of realizing and analyzing polaritonic modes in Mie void cavities naturally arises.

In this paper, we theoretically study the resonant characteristics and polaritonic spectra of spherical cavities loaded (filled) with a generic resonant medium.
We begin with the case of an empty Mie void (spherical cavity) surrounded by a non-dispersive, transparent dielectric medium to establish the dependencies of the quasinormal modes on size and background permittivities.
Next, we examine the polaritonic spectra of these voids loaded with a resonant medium and analyze the coupling of the void QNMs with the medium resonances.
Within this picture, we establish the role of radial mode number and dielectric characteristics on the regime of coupling. 
Finally, we analyze the inverted problem where an empty Mie void is surrounded by a dispersive dielectric medium described by a Lorentz resonant transition. In this problem, we reveal the role of the background dielectric characteristics, mainly losses, on the resulting quality factors of the system and the regime of light-matter coupling. We also identify the regime of spatial QNM localization that occurs due to large absorption losses of the background medium close to the resonance.

\section{Results}

\begin{figure*}[t!]
\centering\includegraphics[width=1.\textwidth]{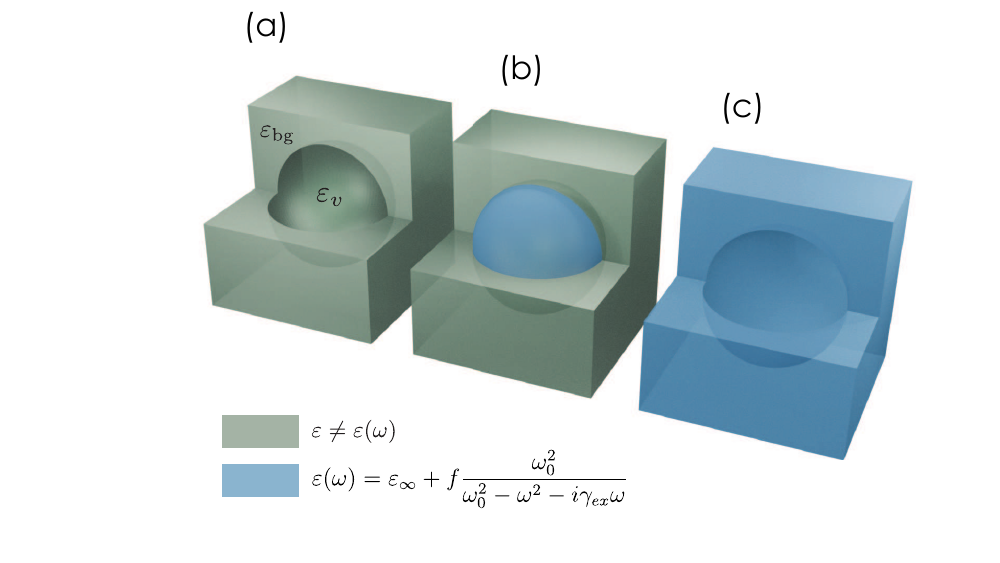}
\caption{Schematics of the three classes of systems analyzed in this paper. 
(a) An empty Mie void inside a transparent dielectric; (b) a Mie void loaded with a resonant medium embedded in a transparent dielectric; (c) an empty Mie void embedded in a dispersive resonant medium. The resonant media are represented by the Lorentz model, Eq.~\eqref{Eq_4}.}
\label{Fig_1}
\end{figure*}

The system studied throughout the paper is represented by a spherical cavity of radius $R$ with permittivity $\eps_v = n_v^2$ surrounded by a background medium $\eps_{\bg} = n_{\bg}^2$.
Both permittivities may be frequency-dependent.
Figure \ref{Fig_1} illustrates three variations of the basic system we analyze in the following: (a) an empty spherical void surrounded by a transparent dielectric background, (b) a dielectric-surrounded spherical void loaded with a resonant medium, and (c) an empty spherical void embedded in a dispersive resonant medium.
This last system establishes a connection and explains certain spectral features observed earlier by Hentschel et al. \cite{hentschel2023dielectric} from the general polaritonic perspective.

The resonant characteristics and polaritonic spectra of Mie voids realized with various materials are encoded in the optical eigenfrequencies of the structures. 
For a spherical cavity, the electromagnetic field of every quasinormal mode (QNM) is represented by a single vector spherical harmonic with a particular polarization state (TE, TM), angular momentum $\ell$, and magnetic number (projection of the orbital momentum) $m$. For any given polarization state and angular momentum $\ell$, complex-valued eigenfrequencies of the spherical cavity are given by the roots of the characteristic equation independent of $m$ \cite{Mie1908}:
\begin{equation}
\begin{split}
    \mathrm{TM:} \quad {n_{\bg} \xi_\ell (n_{\bg} kR) \psi_\ell '(n_v kR) -
    n_v \psi_\ell (n_v kR) \xi_\ell '(n_{\bg} kR)} = 0 ,
\end{split}
\label{Eq_1}
\end{equation}
\begin{equation}
\begin{split}
    \mathrm{TE:} \quad {n_v \xi_\ell (n_{\bg} kR) \psi_\ell '(n_v kR) -
    n_{\bg} \psi_\ell (n_v kR) \xi_\ell '(n_{\bg} kR)}=0
\end{split}
\label{Eq_2}
\end{equation}
Here, $k = \omega/c$, 
$\psi_\ell(x) = x j_\ell(x)$ and $\xi_\ell(x) = x h_\ell^{(1)}(x)$ are Riccati-Bessel functions, 
and $j_\ell(x)$ and $h_\ell^{(1)}(x)$ are spherical Bessel and Hankel functions of the first kind, respectively. The prime denotes derivatives with respect to the argument.
For a non-dispersive real-valued background permittivity, the above characteristic equations yield countably many solutions for any given polarization and $\ell$.
These modes exhibit identical angular but different radial distributions, and are characterized by radial order $N$. We will focus on dipolar ($\ell = 1$) and quadrupolar ($\ell = 2$) modes up to the first four radial orders for a void of fixed radius. By subsequent radius alteration, we will be able to track the eigenfrequencies trajectories on the complex-frequency plane as well as the dispersions associated with the set dielectric characteristics.

\subsection{Spectral properties of empty Mie voids}

\begin{figure*}[t!]
\centering\includegraphics[width=1\textwidth]{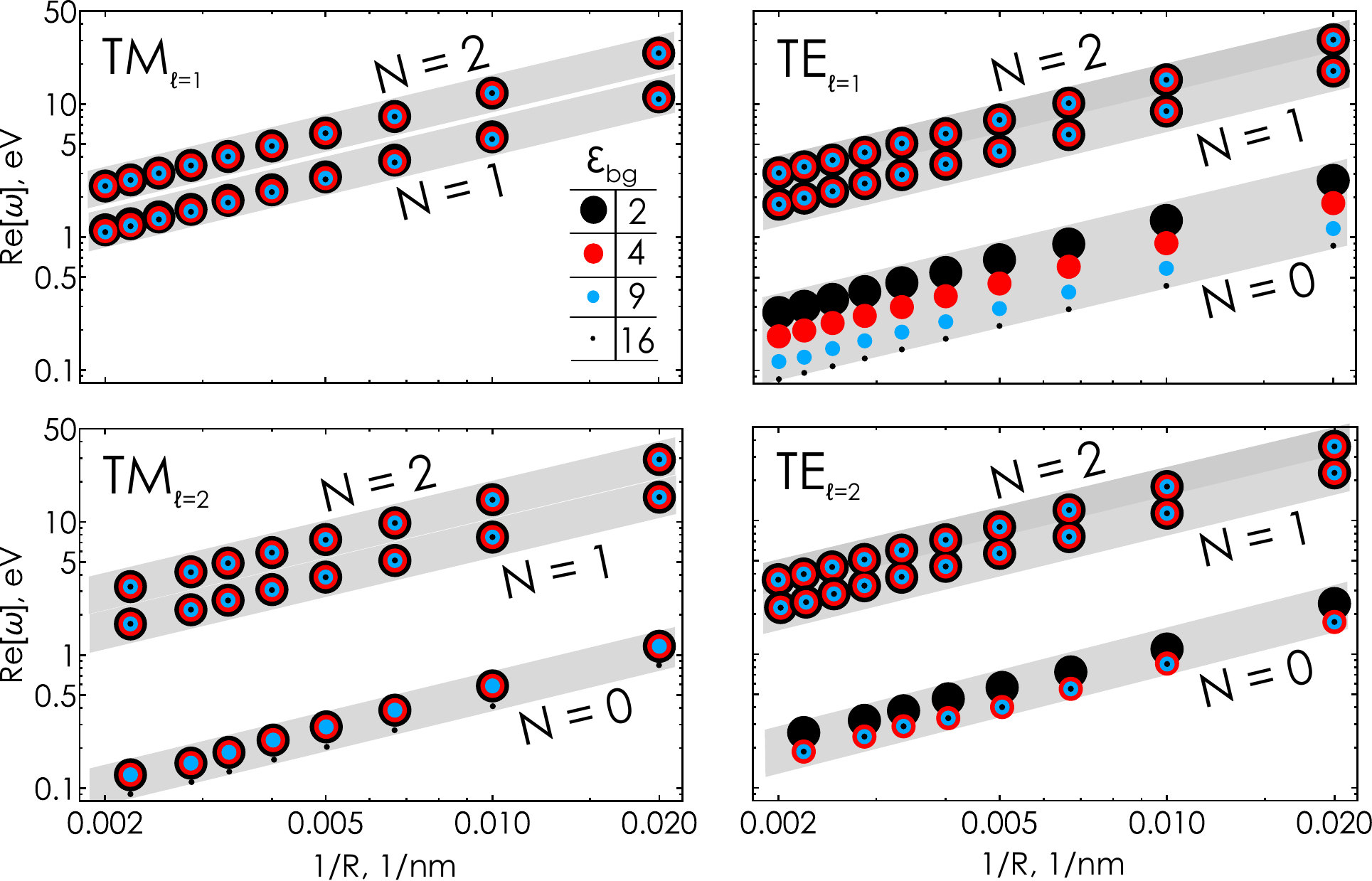}
\caption{Resonant frequencies of the TM$_{\ell=1,2}$ and TE$_{\ell=1,2}$ quasinormal modes of an empty spherical Mie void surrounded by different transparent dielectrics as functions of the void inverse radius in double logarithmic scale. 
Gray areas are guides for the eye, approximately indicating the spectral ranges of each particular mode.
For the electric field profiles of the presented modes, see Fig. S1 of the Supporting Information.
}
\label{Fig_2}
\end{figure*}

We begin by analyzing eigenfrequencies of an empty spherical Mie void ($\mu = 1$, $\eps = 1$) surrounded by a transparent dielectric with refractive index $n_{\bg} = \sqrt{\eps_\bg}$ [see Fig. \ref{Fig_1}(a)].
Figure \ref{Fig_2} shows the resonant frequency (real part of the complex-valued eigenfrequency) as a function of the inverse radius $1/R$ for various permittivities of the background material (different marker colors), polarizations (TM and TE), angular momentum ($\ell = 1,2$), and radial mode numbers $N$. The spectrum of Mie voids consists of two types of modes. The first one corresponds to conventional Mie resonances, which are confined within the void due to the dielectric contrast. The modes of the second type are strongly delocalized, having very low $Q$-factors. The difference in the field distributions for these two types of modes is shown in Fig. S1 of the Supporting Information. In what follows, we will focus on the well-confined Mie modes, numbered from $N=1$. The number of delocalized modes is infinite, but we will account for only the first ones for each polarization and angular momentum and assign to them $N=0$. It is worth mentioning that the entire spectrum also includes static modes with pure imaginary eigenfrequencies (see the Supporting Information Section S3 for details). They provide the completeness of the basis but do not manifest themselves in the scattering spectra explicitly. Figure~\ref{Fig_2} highlights the distinct nature of Mie-void-resonances and delocalized modes. Indeed, the spectral position of Mie-void resonances ($N=1,2$) almost doesn't depend on the background permittivity $\varepsilon_{\rm bg}$ as previously discussed in Ref.~\cite{hentschel2023dielectric}, while the frequency of delocalized modes ($N=0$) are scaled as $1/n_{\rm bg}$ for high values of $n_{\rm bg}$ (see the Supporting Information Section S3). Therefore, the wavelength for such resonances can be substantially larger than the size of the void. However, the use of Mie voids as efficient subwavelength resonators, similar to plasmonic particles, is dubious due to the low $Q$-factors of the delocalized modes.

\begin{figure*}[t!]
\centering\includegraphics[width=1\textwidth]{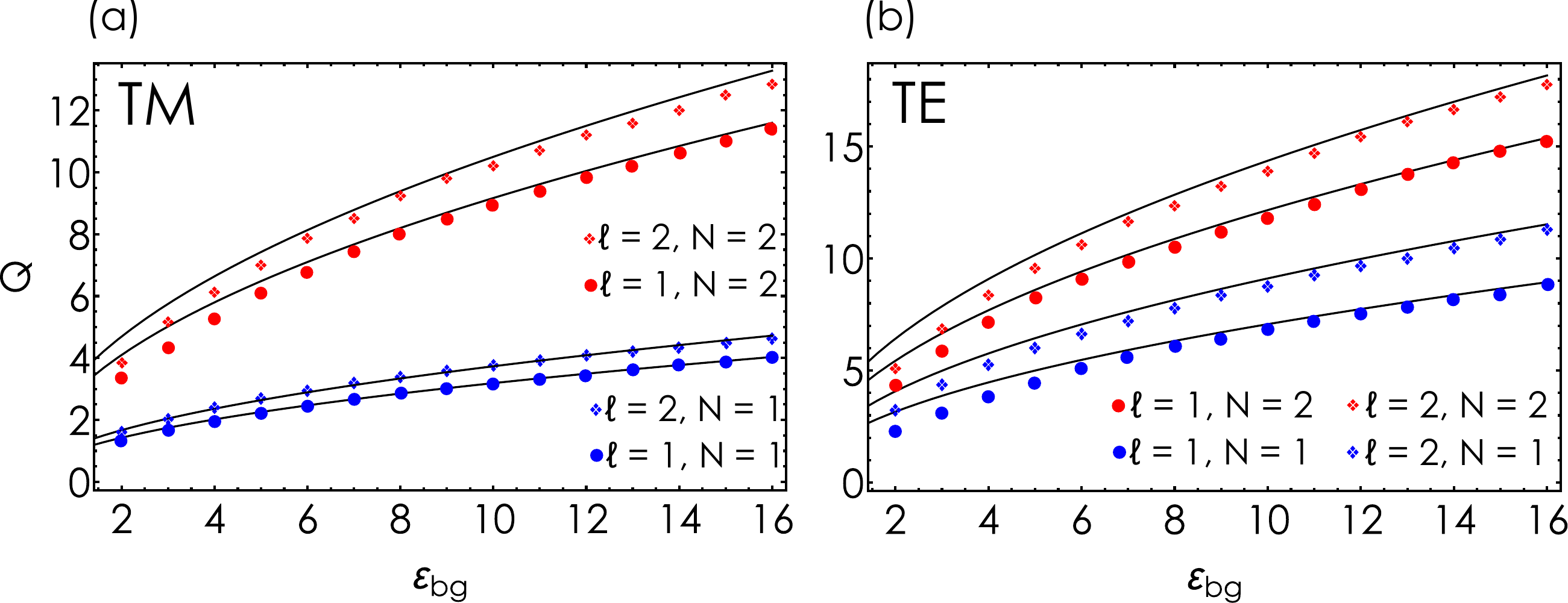}
\caption{$Q$-factors of (a) TM$_{\ell=1,2}$ and (b) TE$_{\ell=1,2}$ modes of a Mie void as functions of the dielectric permittivity of the background material. The black lines represent the analytical expressions for the obtained $Q$-factors, Eq.~\eqref{Eq_3}.
}
\label{Fig_3}
\end{figure*}

The calculated eigenfrequencies $\w$ of empty spherical voids allow us to calculate the quality factors $Q_v = \Re[\omega] / 2 |\Im[\omega]|$.
Figure \ref{Fig_3} shows the resulting $Q$-factors for the dipole and quadrupole Mie-void resonances as functions of the background permittivity. Clearly, the $Q$-factors grow monotonically with increasing background permittivity, in agreement with findings in \cite{hentschel2023dielectric}. The analytical expressions for spectral positions of Mie-void resonances and their $Q$-factors can be found from Eqs.~\eqref{Eq_1} and \eqref{Eq_2} in the limit of high background permittivity as a perturbation series, where $\varepsilon_v/\varepsilon_{\rm bg}\ll1$ plays a role of small parameter (see the Supporting Information Section S2 for details):
\begin{equation}
\begin{split}
    \text{TE modes}: Q_v &\;\approx\; 
    \frac{z_0}{2}\;\sqrt{\frac{\varepsilon_{\rm bg}}{\varepsilon_v}},
    \quad
    j_\ell(z_0)=0,\\[6pt]
    \text{TM modes}: Q_v &\;\approx\; 
    \frac{z_0}{2}\,
    \left|1-\dfrac{\ell(\ell+1)}{z_0^2}\right|\;
    \sqrt{\frac{\varepsilon_{\rm bg}}{\varepsilon_v}},
    \quad  z_0j'_\ell(z_0)+j_\ell(z_0)=0.
\end{split}
\label{Eq_3}
\end{equation}

One can notice that the equations determining the real part of resonant frequencies of TM- and TE-polarized modes match exactly the respective characteristic equations of a spherical PEC cavity \cite{jackson_electodyn}. This is expected, since in the limit $\eps_\bg \to \infty$, the properties of the dielectric background medium approach those of the perfect electric conductor.  The resulting analytical expressions agree well with the numerical results for both TM and TE modes, even for moderately small background permittivities ($\varepsilon_{\rm bg}\gtrsim 2$) (see Fig.~\ref{Fig_3}).

It is interesting to note an extremely slow growth of $Q$-factors of the void modes with $\eps_\bg$ in contrast to much faster growth of $Q$-factors of a spherical dielectric particle in air \cite{ZambranaPuyalto2024}.

\begin{figure*}[t!]
\centering\includegraphics[width=1\textwidth]{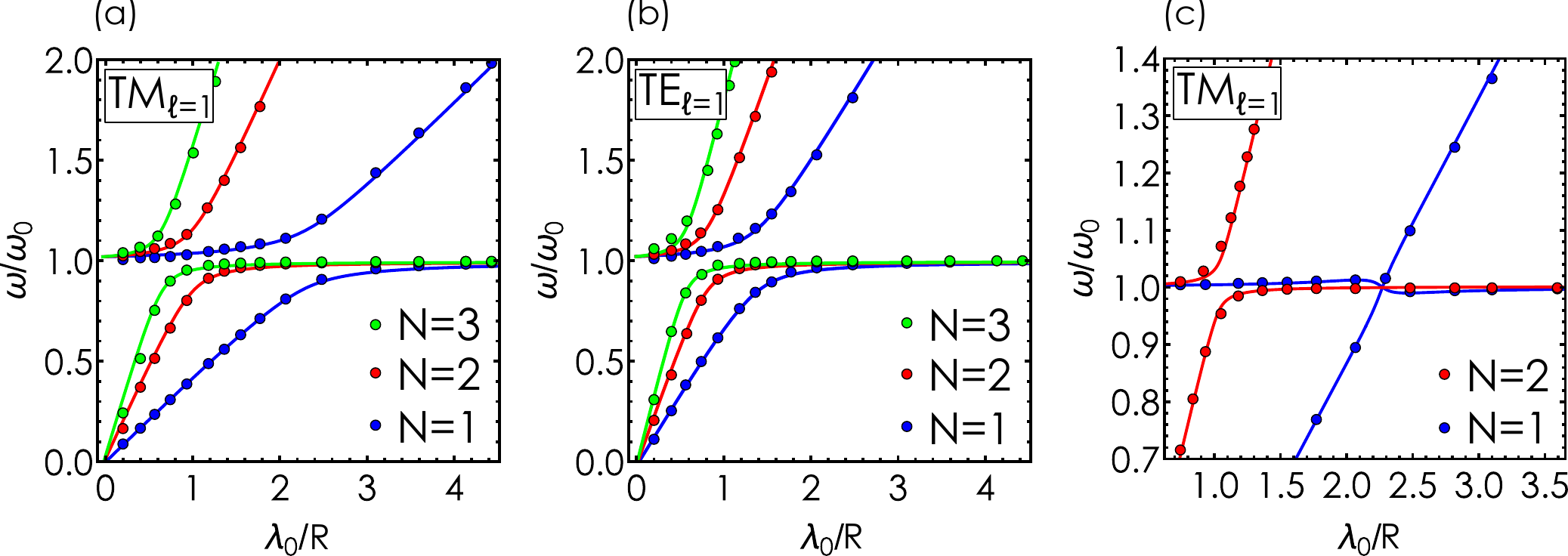}
\caption{Polaritonic spectra of a Mie void surrounded by a transparent dielectric ($\varepsilon_{\bg} = 16$) and loaded with a resonant medium. 
(a), (b) Eigenfrequency spectra of the TM$_{\ell=1}$ and TE$_{\ell=1}$ modes of the dielectric void loaded with a resonant medium ($f = 0.1$, $Q_{ex} = 100$), revealing a set of anti-crossings. 
(c) Same as (a) but for $f = 0.01$, $Q_{ex} = 100$. $\lambda_0 = 2 \pi c/ \w_0$ is the resonant wavelength of the Lorentz material. The colored curves represent the analytical polaritonic spectra, Eq. \eqref{polspec}.}
\label{Fig_4}
\end{figure*}

\subsection{Mie void loaded with a resonant medium}

Next we analyze the polaritonic system [see Fig. \ref{Fig_1}(b)], by filling the spherical void with a resonant medium described by the single-pole Lorentz model:
\begin{equation}
    \eps(\omega) = \varepsilon_\infty + f \frac{\w_0^2}{\w_0^2 - \w^2 - i \gamma_{ex} \w},
    \label{Eq_4}
\end{equation}
where $\varepsilon_\infty$ is the high-frequency permittivity, $f$ is the reduced oscillator strength, $\w_0$ and $\gamma_{ex}$ are the polaritonic resonance frequency and decay rate. 



Substituting Eq.~\eqref{Eq_4} into Eqs.~\eqref{Eq_1} and \eqref{Eq_2} with $\varepsilon_v=\varepsilon(\omega)$, and
linearizing the resulting expressions in the vicinity of
$\omega\approx\omega_v\approx\omega_{0}$, we obtain a compact equation for the
polaritonic spectrum:
\begin{equation}
  \bigl(\omega-\omega_v+i\gamma_v/2\bigr)\,\bigl(\omega-\omega_0+i\gamma_{ex}/2\bigr)
  = \frac{f\,\omega_0^2}{4\varepsilon_\infty}.
  \label{polspec}
\end{equation}
Here, the first bracket corresponds to the geometrically scalable Mie-void resonance, where $\omega_v$ and $\gamma_{v}$ are the Mie-void resonant frequency and decay rate, different for TM- and TE-polarized modes (see Section S4 of the Supporting Information for details). The second bracket describes the bulk excitonic resonance, and the right-hand side encodes coupling between exciton and Mie resonance  with effective coupling constant 
\begin{equation}
g = (\omega_0/2) \sqrt{f/\eps_{\infty}}.
\label{gbulk}
\end{equation}
This expression agrees with those from Ref.~\cite{Canales2021}, confirming that the bulk coupling strength can be taken as a reasonable estimate of the collective coupling strength for spherical resonators loaded with a resonant medium. Equation~\eqref{polspec} is equivalent to the eigenvalue problem of an effective $2\times2$ non-Hermitian Hamiltonian,
\begin{equation}
  \hat H =
  \begin{pmatrix}
    \omega_{0} - i\gamma_{ex}/2 & g \\
    g & \omega_{v} - i\gamma_{v}/2
  \end{pmatrix}.
\end{equation}
The eigenfrequencies of this effective Hamiltonian are given by:
\begin{equation}
\begin{split}
    \w_{\pm} = \frac{\omega_{v} + \omega_0}{2} - \frac{i}{2}\left(\frac{\gamma_{ex}}{2} + \frac{\gamma_{v}}{2}\right) \pm
    \sqrt{g^2 + \frac{1}{4}\left(\omega_{0} - \omega_{v} - i \left(\frac{\gamma_{ex}}{2} - \frac{\gamma_{v}}{2}\right)\right)^2}.
\end{split}
\label{eigfreq}
\end{equation}
The Rabi splitting $\Omega = \w_+ - \w_-$ between the two polaritonic states at zero detuning ($\w_{v} = \w_0$) is real-valued if $4 g>\left| \gamma_{ex} - \gamma_{v}  \right|$.

\begin{figure*}[t!]
\centering\includegraphics[width=1\textwidth]{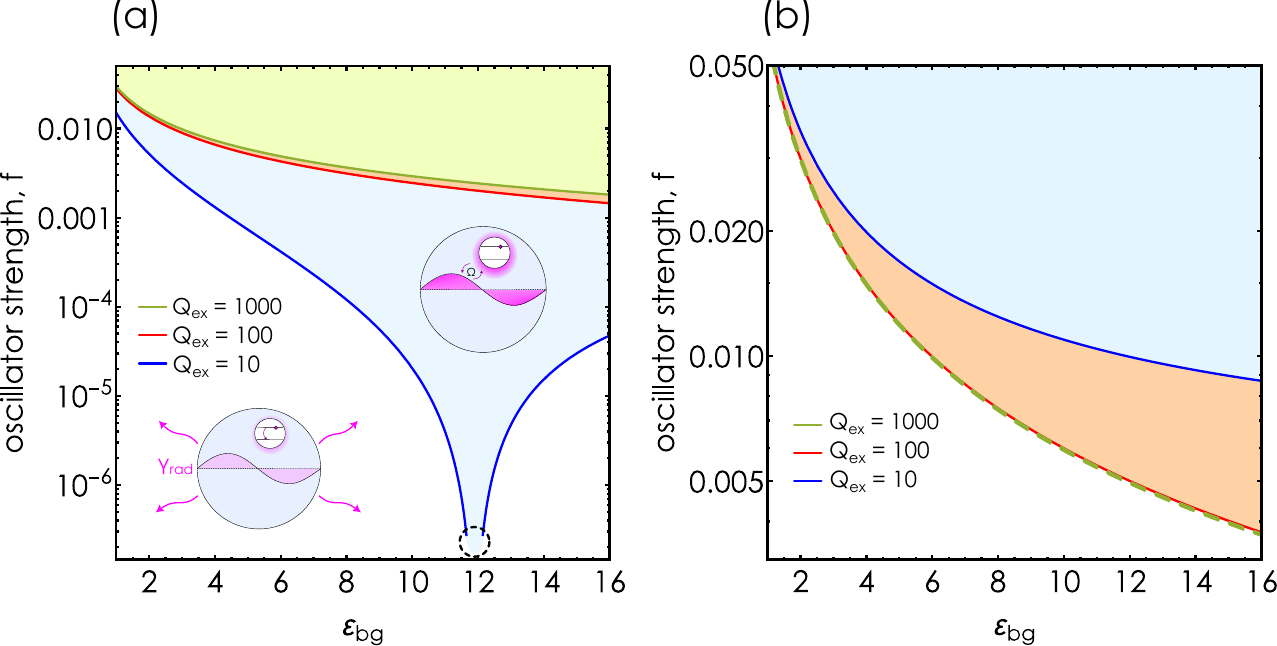}
\caption{
(a) Coupling diagram in the parameter space of the background dielectric permittivity $\eps_\bg$ and reduced oscillator strength $f$ in logarithmic scale for various $Q$-factors of the resonant medium filling the void. The lines indicate the corresponding threshold values of reduced oscillator strength $f_{\mathrm{th}1}$; the colored areas depict strong coupling domains between the TM$_{\ell=1}$ $N=2$ cavity mode and the resonant medium in accordance with Eq. \eqref{Eq_7}. 
Insets are a graphical illustration of weak and strong coupling in Mie voids loaded with resonant media. Black dashed circle marks the point of $Q$-factors matching, $Q_{ex} = Q_{v}$, resulting in strong coupling for arbitrarily low values of $f$.
(b) Same as (a) but for the stronger threshold $f_{\mathrm{th}2}$, Eq. \eqref{Eq_9}.
}
\label{Fig_5}
\end{figure*}

Figures~\ref{Fig_4}(a) and \ref{Fig_4}(b) show the numerically and analytically calculated polaritonic spectra of TM- and TE-polarized dipolar modes of the void  filled with a resonant medium ($f=0.1$, $Q_{ex}=100$) and surrounded by a transparent dielectric ($\varepsilon_{\bg} = 16$). For the sake of simplicity, we set $\eps_\infty = 1$. The analytical spectra [see Eq.~\eqref{polspec}] shown by solid curves are in good agreement with the the numerical results (markers) both on and off excitonic resonance. The spectra clearly exhibit anticrossings, indicating strong coupling between the void mode and the excitonic transition. With increasing radial number $N$, the anticrossing shifts to larger void radii (smaller $\lambda_0/R$), consistent with electromagnetic frequency scaling. Figure~\ref{Fig_4}(c) shows the
corresponding spectra for a weaker oscillator strength ($f=0.01$, $Q_{ex}=100$),
where both weak and strong coupling regimes appear depending on the optical mode.
Each photonic mode generates an independent pair of lower and upper polaritons
\cite{Richter2015}, reflecting the spatially nonuniform exciton--photon coupling.
This contrasts with systems featuring localized, point-like transitions, where
multiple photonic modes couple to a single resonance and produce middle polaritons
\cite{tay2025multimode}.

Expressing the coupling constant as Eq. \eqref{gbulk} in terms of the Lorentz oscillator model, we arrive at the following criterion of observing real-valued Rabi splitting between the two eigenstates of the coupled system (polariton anti-crossing):
\begin{equation}
    f > \left(\frac{1}{2 Q_{ex}} -  \frac{1}{2 Q_{v}} \right)^2,
\label{Eq_7}
\end{equation}
where $Q_{v}$ is the quality factor of the void cavity, 
which can be expressed through $\eps_{\bg}$ by analytical expressions in Eq.~\eqref{Eq_3}. 

Equation \eqref{Eq_7} allows us to introduce the \emph{threshold} value of $f$ delimiting the regions of weak and strong coupling in terms of polaritonic frequency splitting for the given parameters:
\begin{equation}
    f_\mathrm{th1} = \left(\frac{1}{2 Q_{ex}} -  \frac{1}{2 Q_{v}} \right)^2.
\label{f_th}
\end{equation}

Figure \ref{Fig_5}(a) shows the regions of weak (white) and strong (colored) coupling defined according to Eq. \eqref{Eq_7} for the TM-polarized dipolar $N=2$ mode in the parameter space of the background permittivity and the reduced oscillator strength $f$. 
Analytical approximation, Eq. \eqref{Eq_3}, was used to evaluate $Q_{v}$. Higher values of background permittivity favor the emergence of anti-crossing in this system until the point of equal $Q$-factors, $Q_{v} = Q_{ex}$ [black dashed circle in Fig.~\ref{Fig_5}(a)], where polaritonic frequency anti-crossing takes place for arbitrarily low values of $f$, Eq.~\eqref{Eq_7}.

\begin{figure*}[t!]
\centering\includegraphics[width=1\textwidth]{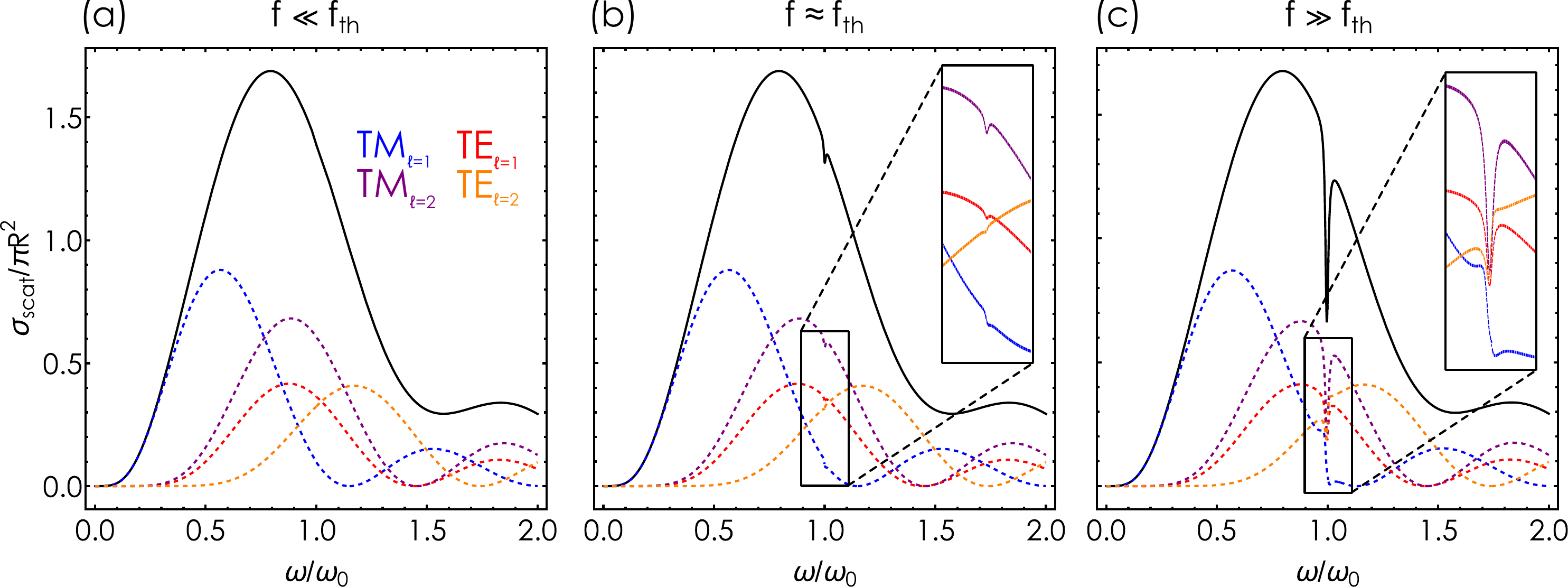}
\caption{The effect of the Lorentz oscillator strength on the scattering cross-section of a Mie void.
(a) Partial and total cross-section of a Mie void for $\eps_\bg = 16$ loaded with a a resonant medium with $Q_{ex}=100$ and $f = 5\cdot10^{-4}$ below the strong coupling threshold ($f_{\mathrm{th}2}$, Eq. \ref{Eq_9}) of the TM$_{\ell =1,2}$ $N = 1$ and TE$_{\ell =1,2}$ $N = 1$ modes.
(b) Same as (a) for $f=5\cdot10^{-3}$ slightly above the strong coupling threshold of the TE$_{\ell =2}$ $N = 1$ mode.
(c) Same as (a) for $f= 0.1$ way above the strong coupling threshold of all the presented modes.
}
\label{Fig_6}
\end{figure*}



On the other hand, in order to observe an anti-crossing in a spectral response of the system (such as extinction cross-section), the system has to follow a more stringent condition of Rabi splitting $\Omega$ exceeding the polariton half-width at half maximum~\cite{Baranov2018}:
\begin{equation}
    \Omega > \frac{\gamma_{ex}}{2} + \frac{\gamma_{v}}{2}.
\label{Eq_8}
\end{equation}
This inequality leads to a stronger criterion for observation of polariton anti-crossing:
\begin{equation}
    f > f_\mathrm{th2}, \quad
    f_\mathrm{th2} =  \left(\frac{1}{2 Q_{ex}^2} +  \frac{1}{2 Q_{v}^2} \right).
\label{Eq_9}
\end{equation}
The respective coupling diagrams for three different resonant media quality factors $Q_{ex}$ are presented in Fig. \ref{Fig_5}(b). 
For this system, higher values of $\eps_{\bg}$ guarantee lower values of $f_{th2}$.
The two criteria tend to show identical coupling regions in the limit of high $Q_{ex}$ and $Q_{v}$.


In addition, Fig. \ref{Fig_6} shows the evolution of partial and total scattering cross-sections of a Mie void loaded with a Lorentz medium in different coupling regimes.
For $f$ below the strong coupling threshold, Fig. \ref{Fig_6}(a), cross-sections do not show any signs of light-matter coupling, as expected.
Exactly at the strong coupling threshold, $f = f_{\mathrm{th}2}$, all partial and total cross-sections start to show signs of coherent light-matter coupling, as evidenced by the development of a Fano resonance around the exciton frequency $\w_0$, Fig. \ref{Fig_6}(b) \cite{Hartsfield2015, zhang2017sub, wang2017coherent,pantyukhina2025excitonic}.
Finally, deep in the strong coupling regime, $f \gg f_{\mathrm{th}2}$, the Fano resonance deepens in all partial cross-sections; nonetheless, clear mode splitting is hard to identify even in this regime.
This is typical/characteristic for multi-mode structures: even partial cross-section for a fixed polarization and angular momentum accounts for multiple FP-like Mie modes, whose presence often does not allow observing a clear anti-crossing in extinction or scattering spectra \cite{canales2024self}.

\subsection{Empty Mie void surrounded by a dispersive medium}

Finally, we examine the eigenfrequency spectra of an empty spherical void ($\eps_v = 1$) embedded in a dispersive resonant medium, Fig. \ref{Fig_1}(c), described by the Lorentz model, Eq. \eqref{Eq_4}.  
To obtain the practical range of oscillator strength values of the high-refractive index background medium, we fit the complex-valued permittivity of crystalline silicon from \cite{Aspnes1983} with a set of Lorentz poles:
\begin{equation}
    \eps(\omega) = \varepsilon_\infty + \sum_n{f_n \frac{\w_n^2}{\w_n^2 - \w^2 - i \gamma_{n} \w}}.
    \label{Eq_13}
\end{equation}
The resulting fit is shown in Fig. S2 with the corresponding reduced oscillator strength values. 
According to these values, we extend the range of the oscillator strength beyond $f = 1$ in the single-pole model [Eq. \eqref{Eq_4}], and set $\eps_\infty = 16$ for the following calculations.

\begin{figure*}[t!]
\centering\includegraphics[width=1\textwidth]{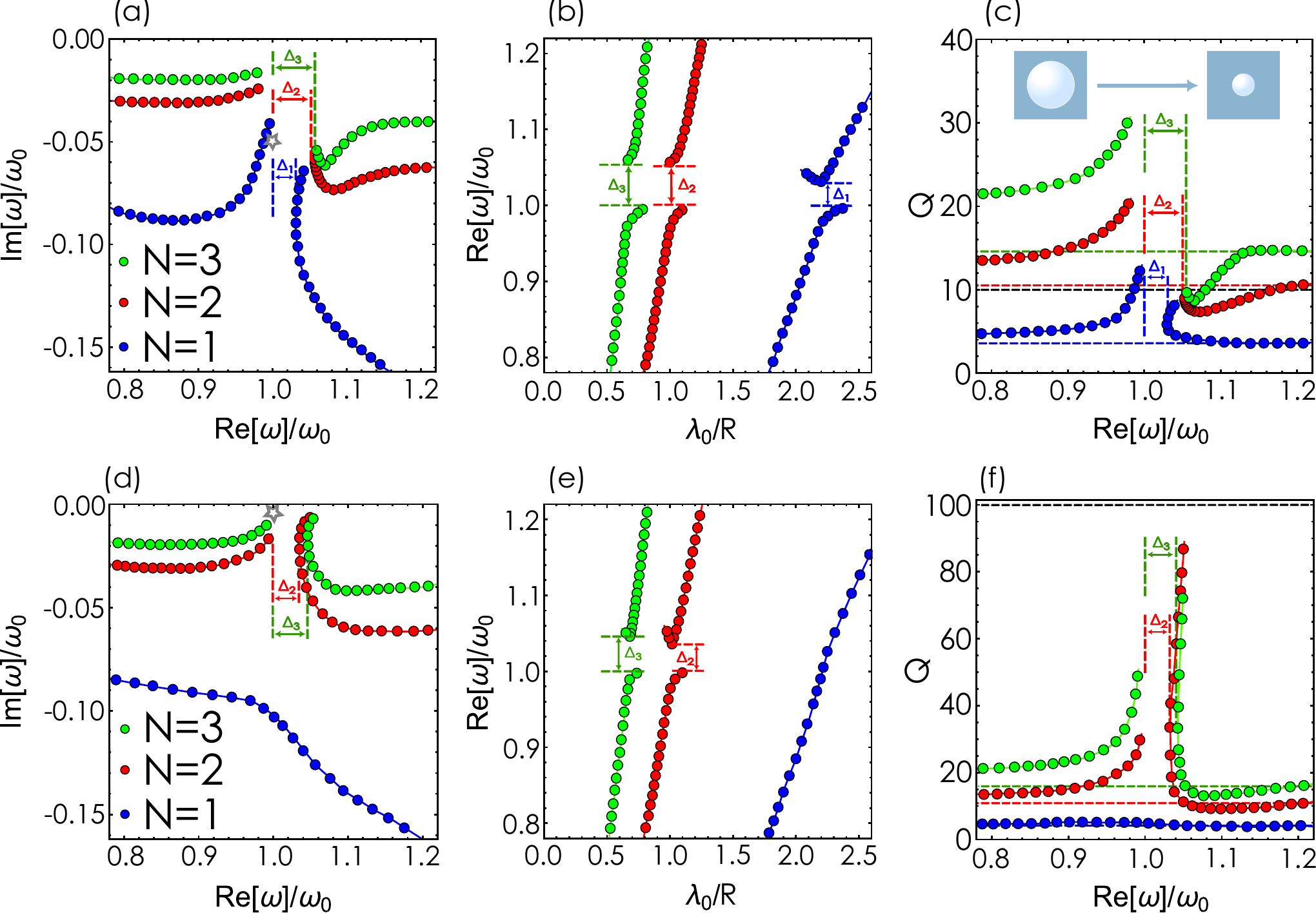}
\caption{
The effect of the background resonant medium on the behavior of eigenfrequencies and $Q$-factors of a Mie void. The results were obtained for the TM$_{\ell=1}$ mode, low-$Q$ (top row, $Q_{ex} = 10$) and high-$Q$ (bottom row, $Q_{ex}=100$) resonant background media with $f = 2$ and $\varepsilon_{\infty} = 16$. 
(a) Trajectories of polaritonic eigenfrequencies in the complex frequency plane for the $N = 1, 2, 3$ modes; the gray star indicates the background medium resonant frequency.
$\Delta_i$ marks the position of a respective polariton gap for each mode.
(b) Real parts of eigenfrequencies from (a) plotted as a function of the normalized inverse radius of the Mie void.
(c) $Q$-factors of polaritonic modes as functions of the normalized resonant frequency. 
The horizontal black and colored dashed lines mark the quality factors of the background medium and photonic modes, respectively.
(d) - (f) Same as (a) - (c) but evaluated for the case of a high-$Q$ ($Q_{ex}=100$) background medium. Inset in (c) shows the direction of the void radius sweep. The thick colored curves are guides for the eye connecting the numerically obtained values.
}
\label{Fig_7}
\end{figure*}

Figure \ref{Fig_7} shows the behavior of complex-valued eigenfrequencies of an empty spherical void in a resonant background for the cases of low-$Q$, $Q_{ex} \approx Q_{v}$ (top row), and high-$Q$, $Q_{ex} \gg Q_{v}$ (bottom row) resonant media.
The results were obtained for the first three radial numbers $N = 1,2,3$ of a TM-polarized $\ell = 1$ mode. 
It is instructive to analyze the eigenfrequency trajectories in the complex-frequency plane, Fig. \ref{Fig_7}(a,d). 
For some sets of parameters, the eigenfrequencies demonstrate splitting into upper (UP) and lower (LP) polariton branches in the vicinity of the excitonic complex frequency $\w_0 - i \gamma_{ex}/2$ (gray stars).
The pairs of LP and UP that exhibit anti-crossing additionally feature the formation of polaritonic gaps, $\Delta$, except for the TM-polarized $\ell = 1$, $N=1$ mode, Fig. \ref{Fig_7}(d). 
The gap generally widens with increasing radial number $N$.
The next column, Fig. \ref{Fig_7}(b,e), for convenience displays the same set of data plotted in the coordinates of normalized resonant frequency $\Re[\omega] / \w_0$ and normalized inverse radius $\lambda_0 / R$, where $\lambda_0 = 2 \pi c/ \w_0$ is the resonant wavelength of the Lorentz material.

In systems possessing translational invariance, the upper edge of the polariton gap can be calculated as the energy of the upper polariton in the limit of $k\to0$, 
$\bar{\w} = \sqrt{\w_0^2 + 4g^2}$ \cite{Todorov2012}. 
Conversely, the lower edge can be calculated as the energy of the lower polariton in the limit of $k\to\infty$, which is exactly the uncoupled excitonic resonant frequency $\w_0$ \cite{Todorov2012}.
In compact polaritonic structures -- such as the one studied here -- the in-plane wave-vector component can be replaced with the inverse characteristic dimension of the system, $k_\parallel \to 1/R$.
Figures \ref{Fig_7} (b,e) show that the lower edge of the polariton gap indeed coincides with the frequency of the uncoupled resonant oscillations of the background medium.
However, the upper edge of the gap does not occur at $r\to \infty$, and instead occurs at intermediate values of $r$.
Furthermore, the spectra of the upper polariton display a characteristic back-bending picture [see $N=1$ mode in Fig. \ref{Fig_7}(b) and $N = 2,3$ modes in Fig. \ref{Fig_7}(e)].

Next, we analyze this data from the perspective of the $Q$-factors of the resulting modes, Fig. \ref{Fig_7}(c,f).
The observed splitting is accompanied by an increase in the quality factor of the resulting polaritonic modes in comparison with the uncoupled photonic void modes.
We note that the behavior of polaritonic $Q$-factors is qualitatively different depending on the ratio between $Q_{v}$ and $Q_{ex}$.
In the case of low-$Q$ background medium, $\gamma_{ex} \approx \gamma_{v}$, the $Q$-factor of the lower polariton mode exceeds both the photonic \emph{and} the excitonic $Q$-factors, Fig. \ref{Fig_7}(c). 
This behavior is similar to the observation of polariton linewidth asymmetry in structures incorporating plasmonic meta-atoms \cite{canales2023polaritonic}.
In the case of high-$Q$ background, $\gamma_{v} \gg \gamma_{ex}$, the $Q$-factors of both polariton modes exceed the bare cavity $Q$-factor only and approach that of the exciton, Fig. \ref{Fig_7}(f).

This kind of $Q$-factor improvement has also been discussed by Hentschel et al. \cite{hentschel2023dielectric}. 
The authors explained this effect by the greater refractive index mismatch in the region of absorption lines of the background medium. 
Our polaritonic picture of light-matter interaction in this class of systems offers an alternative partial interpretation of this behavior.

To illustrate the polaritonic nature of the quality factor improvement, we analyze the $Q$-factor dynamics from the perspective of the coupled oscillators model.
To that end, we show in  Fig. S3 the $Q$-factors obtained from Eq. \eqref{eigfreq} for the same set of cavity mode and Lorentz resonance parameters; as a reasonable estimation, the value of the coupling strength was determined by Eq. \eqref{gbulk}.
The results of the $2\times2$ Hamiltonian model do qualitatively agree with the exact Mie solution in the case of high-$Q$ resonant background, $\gamma_{ex} \ll \gamma_{v}$  (see Fig. S3(d, e, f)).
Within the $2\times2$ model, the two polaritons equalize their decay rates at the zero detuning: $\gamma_{UP} = \gamma_{LP} = (\gamma_{v} + \gamma_{ex})/2$ (see Eq. \eqref{eigfreq}). 
Under the condition $\gamma_{ex} \ll \gamma_{cav}$, this causes both polaritonic modes to acquire lower decay rates compared to the uncoupled cavity mode, $\gamma_{LP, UP} \approx \gamma_{ex}/2$.
Close to the polariton gap, however, the $Q$-factors increase even further beyond this simple factor-2 improvement. 
Indeed, close to the gap, the upper and lower polariton modes become purely excitonic, and acquire the corresponding $Q$-factor, $Q_{UP} \approx Q_{LP} \approx Q_{ex}.$

On the other hand, there is almost no correspondence between the exact Mie solution and the $2\times2$ model in the case of low-$Q$ background medium, $\gamma_{ex} \approx \gamma_{v}$ (see Fig. S3(a, b, c)).
We attribute this to the strongly inhomogeneous field distribution of the quasinormal modes of the spherical voids, which leads to space-dependent coupling strength $g$ and invalidates the simple $2\times2$ model.
Another probable explanation of this behavior lies in the real-valued nature of the coupling constant assumed in the $2\times2$ model.
The radiative line narrowing is known to occur in resonant systems, where a few optical modes interfere destructively via the scattering channel. This is commonly described by imaginary part of the coupling constant \cite{Hsu2016, Bogdanov2019,kivshar2023bound}.
This via-the-continuum interaction typically leads to the formation of a super~\cite{CanosValero2025} and  sub-radiant modes, commonly referred to as the quasi-BIC \cite{Friedrich1985, Rybin2017, Bogdanov2019, Koshelev2019, KoshelevBogdanovBIC2018}. The imaginary part of the coupling constant can be introduced phenomenologically or via more accurate expansion of Eqs.~\eqref{Eq_1} and \eqref{Eq_2} up to higher-order terms.

\begin{figure*}[t!]
\centering\includegraphics[width=1\textwidth]{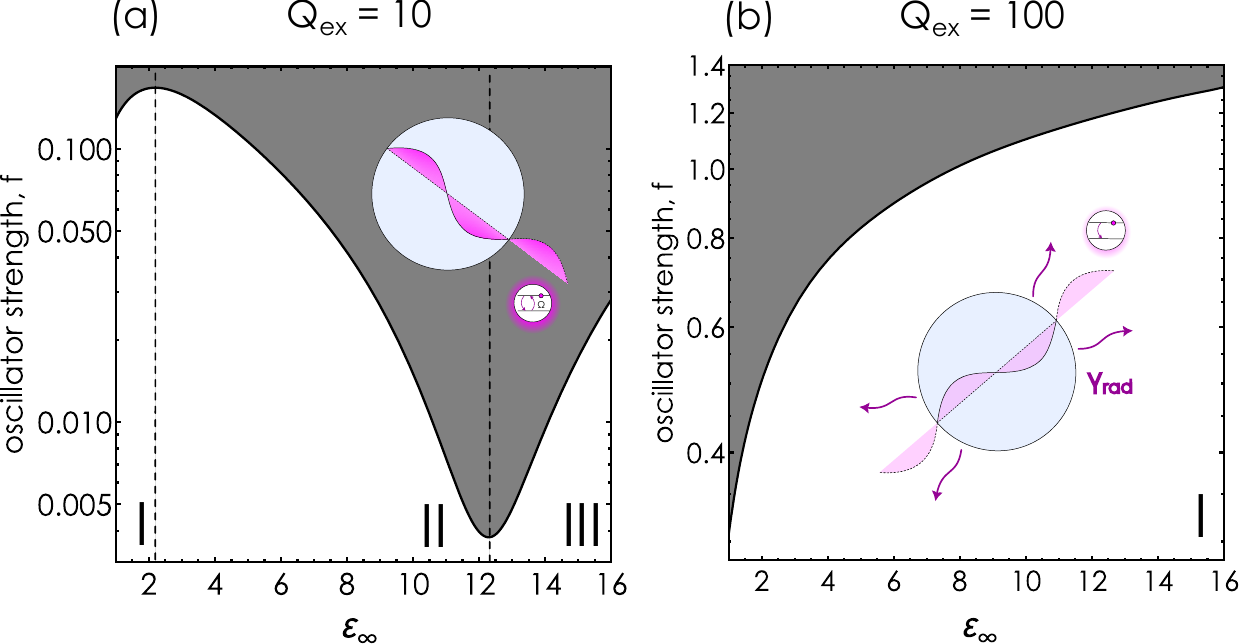}
\caption{Coupling diagrams in the parameter space of the high-frequency background permittivity $\eps_{\infty}$ and reduced oscillator strength $f$ in logarithmic scale. The gray and white areas show the numerically resolved strong and weak coupling regimes between the TM$_{\ell=1}$ $N=2$ cavity mode and the (a)low-$Q$ and (b)high-$Q$ resonant background media. Insets: free interpretation of weak and strong coupling in Mie voids embedded in resonant medium.
}
\label{Fig_8}
\end{figure*}

Similarly to the analysis in Section 2.2, we plot the regions of weak and strong coupling in terms of polaritonic frequencies anti-crossing in the $\eps_{\infty}$-$f$ parameter space for the TM-polarized $\ell = 1$, $N=2$ mode 
and 2 different quality factors of the background medium, $Q_{ex} = 10$ and $Q_{ex} = 100$, Fig. \ref{Fig_8}.
The resulting diagrams display a more complicated behavior compared to those for a cavity loaded with resonant media in a transparent background.
The transition between regions II and III (where the strong coupling region meets the $f$ = 0 axis) is analogous to that in medium-loaded cavities: increasing $\eps_\infty$ improves the photonic mode $Q$-factor and favors the emergence of polaritonic modes, thus reducing threshold value of $f$, until the system reaches the point of equal $Q$-factors, where the $f_\mathrm{th}$ tends to 0, Fig. \ref{Fig_8}(a).

However, in contrast to the results presented in Fig. \ref{Fig_5}, here we observe an unexpected transition to region I at low $\eps_\infty$, where $f_\mathrm{th}$ again grows with $\eps_\infty$.
In contrast to the empty Mie void surrounded by a transparent background, the resonant background medium here interacts with the exponentially diverging tails of the quasinormal mode.
On one hand, low $\eps_{\infty}$ deteriorates the photonic $Q$-factors and should intuitively lead to increasing $f_\mathrm{th}$.
However, the fields of the mode itself become even more delocalized and penetrate further into the background where the interaction with the resonant medium occurs.
This argument could explain the observed behavior of the threshold value of oscillator strength in region I.
In the case of the resonant medium with a higher $Q$-factor, Fig. \ref{Fig_8}(b), we observe a part of region I accompanied by growing $f_{th}$ with $\eps_\infty$. 
Transition to regions II and III in this case also does occur, but at much greater $\eps_\infty$ yielding appropriate photonic $Q$s.

\begin{figure*}[t!]
\centering\includegraphics[width=1\textwidth]{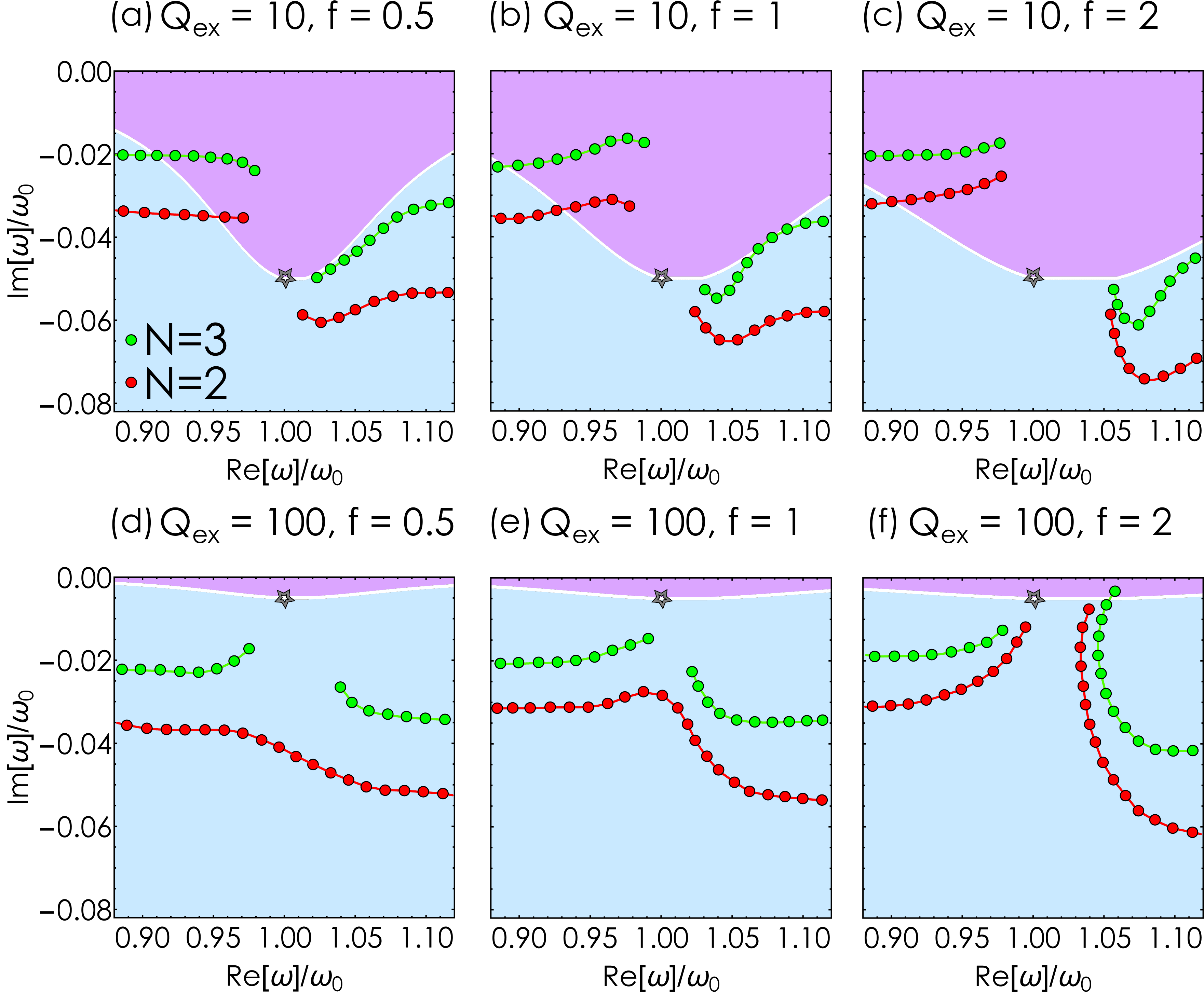}
\caption{QNM localization diagrams in the normalized complex-frequency parametric space for a set of resonant background quality factors and oscillator strengths, $\eps_{\infty}$ = 16. 
Magenta (cyan) areas indicate where the solutions present localized (delocalized) fields.
The gray stars represent the positions of the medium resonances, and the overlayed polaritonic eigenfrequency trajectories represent the TM$_{\ell=1}$ $N=2,3$ modes with the thick colored curves to lead the eye.
}
\label{Fig_9}
\end{figure*}

The presence of frequency dispersion in the background affects not only the spectrum of polaritonic frequencies, but also the spatial behavior of the resulting QNMs.
Typical problems involving dielectric spherical particles imply a transparent background medium, resulting in exponentially diverging fields in space \cite{Lalanne2018}.
The radial part of the QNM field supported by the spherical cavity is described by the spherical Hankel functions of the first kind, $h^{(1)}_\ell (kr)$, which in the far-field zone asymptotically behave as
\begin{equation}
    h^{(1)}_\ell (kr) \to (-i)^{\ell+1} \frac{e^{ikr}}{kr}.
\label{Eq_10}
\end{equation}
Absorption in the background medium may result in spatial localization of the QNM fields.
Specifically, localization occurs when $\Im[k]>0$:
\begin{equation}
    \Im \Biggl[ (\w' + i \w'') 
    \sqrt{ \eps_\infty + \frac{f \omega_0^2}{\omega_0^2 - (\w' + i \w'')^2 - i\gamma_{ex}(\w' + i \w'')} }  \Biggr]  > 0.
\label{Eq_11}
\end{equation}
This expression clearly highlights the competition between two factors determining the possible localization of the QNM: negative imaginary part of the QNM eigenfrequency $\w'' < 0$ leading to a spatially diverging behavior, and the positive imaginary part of the refractive index of the Lorentz medium $\Im \sqrt{\eps} > 0$, leading to spatial localization.

\begin{figure*}[t!]
\centering\includegraphics[width=1\textwidth]{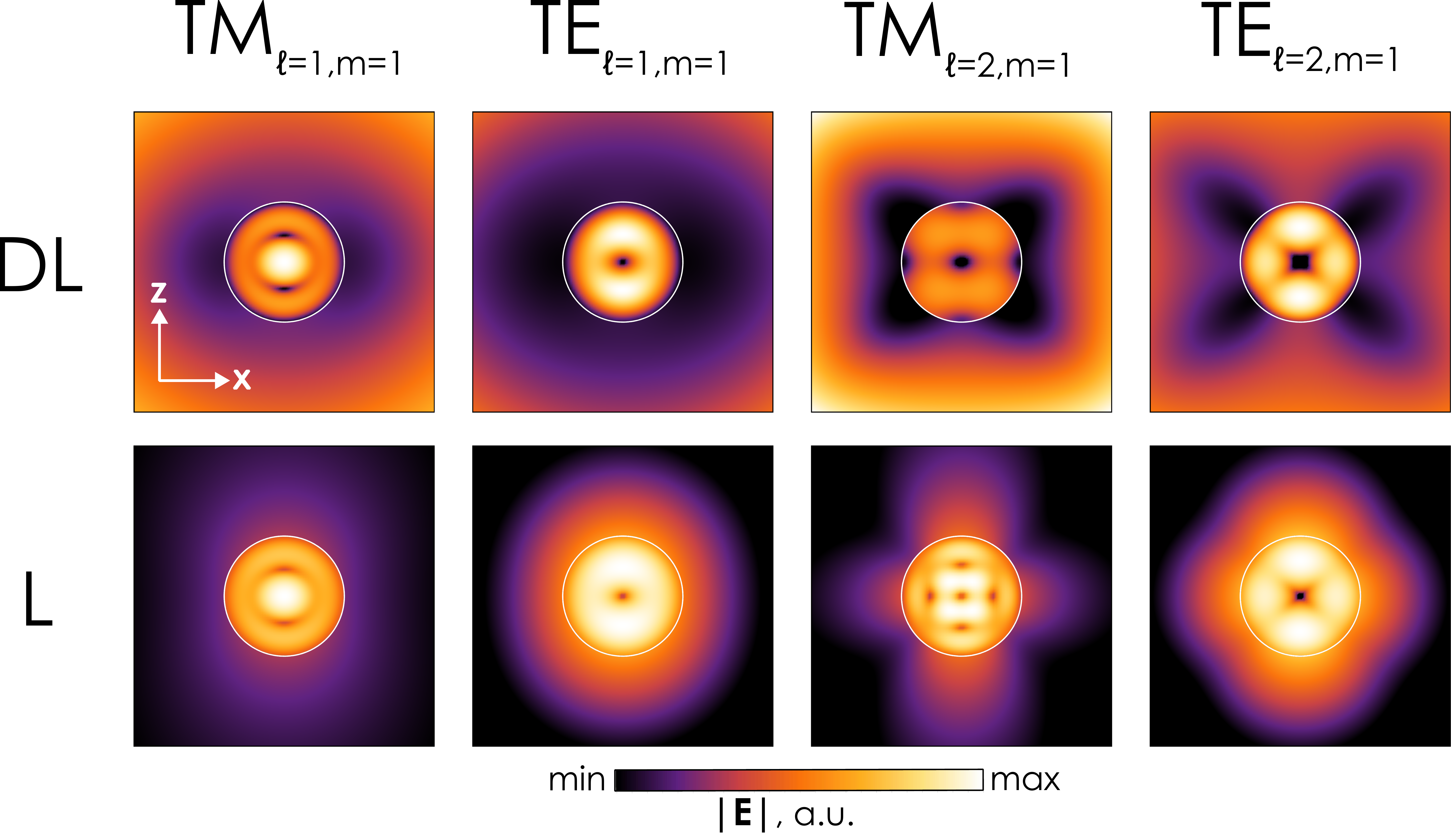}
\caption{Field profiles for delocalized (DL, top row) and localized (L, bottom row) quasinormal modes of Mie voids surrounded by a dispersive dielectric background, Eq. \eqref{Eq_4}. Table S1 of SI contains the numerical values of frequency and medium parameters used for plotting the field distribution of each particular quasinormal mode.
}
\label{Fig_10}
\end{figure*}

Figure \ref{Fig_9} shows the regions of QNM localization in the complex-frequency plane for a set of the resonant medium oscillator strengths $f$ and the exciton quality factors, defined by Eq. \eqref{Eq_11}. 
Clearly, larger values of the oscillator strength $f$ and decay rate $\gamma_{ex}$ of the background medium favor QNM localization. 
The localization regions do not extend beyond $-\gamma_{ex}/2$ (gray stars) along the imaginary frequency axis. 
Each panel of Fig. \ref{Fig_9} additionally demonstrates the trajectories of polaritonic eigenfrequencies for TM-polarized $\ell = 1$, $N = 2,3$ modes overlayed on top of the localization regions. 
Panels (a,b,c) of Fig. \ref{Fig_9} show that the solutions on the lower polariton branch in the case of the low-$Q$ resonant medium regime enter the localization region in the vicinity of resonant frequency in the limit of $r \to 0$. 
This, however, does not occur in the opposite high-$Q$ resonant medium regime, Fig. \ref{Fig_9}(d, e, f).
Generally, for such a polaritonic mode to become localized, one needs to combine a high-$Q$ photonic mode with a low-$Q$ resonant medium.

Finally, we present in Fig. \ref{Fig_10} spatial electric field cross-sections for a few selected QNMs of an empty spherical void in a resonant absorbing background demonstrating either typical delocalized (DL) or localized (L) spatial behavior for the sake of illustration. 
The presented localized modes correspond to the lower-polariton solutions resulting from Mie theory, with eigenfrequencies consistent with the condition presented by Eq. \eqref{Eq_11}. 
This clearly shows that polariton anti-crossing and $Q$-factor enhancement resulting from strong coupling between the cavity modes of a Mie void and the background material are partially responsible for spatial field localization for certain sets of parameters. The exact frequency and medium parameter values utilized in Fig.~\ref{Fig_10} are presented in Table S1 of the Supporting Information.

\section{Conclusion}

To conclude, we have investigated the spectral properties of optical Mie voids in three distinct scenarios: (i) empty voids in transparent dielectrics, (ii) voids loaded with resonant media, and (iii) empty voids embedded in dispersive resonant media. The eigenfrequencies of Mie voids in nondispersive media are almost independent of the host refractive index as long as it is higher than that of the voids. The $Q$-factors of Mie voids (in fact, of generic high-refractive-index-contrast spherical cavities) obey analytical equations similar to those of spherical PEC cavities.

When the void is filled with a resonant medium, the system exhibits both weak- and strong-coupling regimes depending on the radial order of the mode, as confirmed analytically and by numerical simulations. Generally, a higher refractive index contrast at the interface between the void loaded with resonant media and the transparent host medium makes the system less demanding for reaching strong coupling. 

In the case of empty Mie voids embedded in a dispersive resonant medium, the quasinormal modes penetrating the host material demonstrate the polariton anti-crossing with the formation of polariton gaps generally increasing with the radial mode number. The interaction between the void eigenmodes and the host media resonances leads to the enhancement of the polariton $Q$-factors, even exceeding those of the bare components in the case of matching individual $Q$'s, probably owing to the via-the-continuum nature of their interaction. Moreover, the combination of absorption in the host medium and the $Q$-enhanced polariton anti-crossing can lead to the QNM localization governed by the complex eigenfrequencies, as well as the material parameters. This effect can be used for spatially localizing the modes that would not have been localized otherwise for certain systems.

Our theoretical framework establishes Mie voids as a versatile platform for polaritonic engineering, offering distinct advantages in high-permittivity environments. The scaling of $Q$-factors with background permittivity provides a natural route to suppress radiative losses, while the onset of strong coupling positions Mie voids as a prospective platform for polaritonic applications. Our results highlight the potential of Mie voids in all-dielectric nanophotonics, where conventional cavity designs face fundamental limitations.

\section{Acknowledgements}
We thank Kirill Koshelev for stimulating discussion.
The work was supported by the Ministry of Science and Higher Education of the Russian Federation (FSMG-2024-0014).
D.G.B. and E.R. acknowledge support from Russian Science Foundation (grant No. 23-72-10005).
The work was supported by the Ministry of Science and Higher Education of the Russian Federation (FSMG-2024-0014).
D.G.B. and E.R. acknowledge support from Russian Science Foundation (grant No. 23-72-10005). A.A.B. acknowledges support from the National Natural Science Foundation of China (Project W2532010).

\bibliography{mievoid}

\end{document}